\documentclass[10pt,emptycopyrightspace]{ewsn-proc}

\usepackage{xspace}
\usepackage{balance}
\usepackage{graphicx}
\usepackage[hyphens]{url}
\usepackage{epstopdf}
\usepackage{microtype}
\usepackage{booktabs}
\usepackage[table,xcdraw]{xcolor}
\usepackage{caption}
\usepackage{subcaption}
\usepackage{amsmath}
\usepackage{enumitem}
\usepackage{amssymb}
\usepackage{pifont}
\usepackage{algorithm}
\usepackage{algorithmic}

\newcolumntype{P}[1]{>{\centering\arraybackslash}p{#1}}
\newcolumntype{M}[1]{>{\centering\arraybackslash}m{#1}}

\newcommand{\fakepar}[1]{\smallbreak\noindent{#1}}
\newcommand{\boldpar}[1]{\smallbreak\noindent\textbf{#1.}}
\newcommand{\italicpar}[1]{\smallbreak\noindent\textit{#1.}}

\iftrue
    \newcommand{\todo}[1]{\color{red}TODO: #1\color{black}\xspace}

\else
    \newcommand{\todo}[1]{}

\fi
\newcommand{\ieee}{\mbox{IEEE~802.15.4}\xspace}

\newcommand{\blefive}{\mbox{BLE\,5}\xspace}
\newcommand{\dcubee}{\mbox{D-Cube}\xspace}

\newcommand{\NAME}{\mbox{\texttt{OSF}}\xspace}

\usepackage{manyfoot}
\DeclareNewFootnote{A}
\DeclareNewFootnote{B}

\let\footnoteR\footnoteB
\let\footnote\footnoteA
\iftrue
    \newcommand{\michael}[1]{\footnoteR{{\color{red}\bf MB: #1}\color{red}}}
\else
    \newcommand{\michael}[1]{}
\fi

\begin{document}

\title{
    \NAME: An Open-Source Framework for Synchronous Flooding \\over Multiple Physical Layers
}

\iffalse
\author{
    \alignauthor Anonymous Author(s) \\
	 EWSN 2022 -- Submitted Paper \#23
}
\else
\author{
\alignauthorpage Michael Baddeley$^{\ast}$,~Yevgen Gyl$^{\dagger}$,~Markus Schu{\ss}$^{\mathsection}$,~Xiaoyuan Ma$^{\ddagger}$,~and~Carlo Alberto Boano$^{\mathsection}$
    \\
    \affaddr{$^{\ast}$Technology Innovation Institute, AE;\,\,\,
             $^{\dagger}$Unikie, FI;\,\,\,
             $^{\ddagger}$SKF Group, CN;\,\,\,
             $^{\mathsection}$Graz University of Technology, AT} \vspace{1mm} \\
    \email{michael.baddeley@tii.ae}, \email{yevgen.gyl@unikie.com}, \vspace{-3mm} \\
    \email{ma.xiaoyuan.mail@gmail.com}, \email{\{markus.schuss; cboano\}@tugraz.at}
    \vspace{-2.00mm}
}
\fi

\maketitle

\begin{abstract}
Flooding protocols based on concurrent \mbox{transmissions are} regarded as the most reliable way \mbox{to collect or disseminate} data across a multi-hop low-power wireless mesh network. 
Recent works have shown that such protocols are effective for narrowband communication not only over \ieee, but also over the \blefive physical layers (PHYs). 
However, to date, existing literature has only built synchronous flooding solutions on top of a \emph{single} PHY, and there has been no attempt to leverage \emph{different} PHYs \emph{at runtime} \mbox{to increase performance}. 
This paper fills this gap and presents \NAME, an open-source framework that enables the design of multi-PHY synchronous flooding solutions thanks to a novel radio driver and middle-ware architecture capable of dynamically switching the underlying physical layer. 
This allows exploitation of the specific benefits of each PHY (e.g., higher data-rate, increased robustness) on-demand during each flood, increasing performance. 
We tailor \NAME to the off-the-shelf \texttt{nRF52840} platform, and showcase its benefits by comparing single-PHY and multi-PHY synchronous flooding solutions on a real-world testbed. 
\end{abstract}

\iftrue
\fi

\section{Introduction}
\label{sec:introduction}
Interest in synchronous flooding (SF) solutions for low-power wireless networks has gained considerable interest since the observation that collisions in the presence of concurrent transmissions (CT) are not necessarily harmful~\cite{zimmerling2020synchronous}. 
\mbox{The research} community has especially showcased the benefits of SF for narrowband communication, with a large number of data collection and dissemination schemes developed on top of \ieee devices: these have been empirically shown to outperform conventional multi-hop routing approaches across a number of key dependability metrics~\cite{boano17competition, schuss2017competition}. 

With the increasing number of off-the-shelf hardware platforms offering support for multiple narrowband PHYs\footnote{Examples are the Texas Instruments \texttt{CC2652R} / \texttt{CC1352} and the Nordic Semiconductor \texttt{nRF52}~/~\texttt{nRF53} families of low-power wireless devices, which support both \ieee and the four \blefive physical layer modes.}, \mbox{researchers} have shown that SF is also effective over the four physical layers supported by \blefive~\cite{alnahas2019concurrentBLE5}, but emphasized that the PHY selection has a strong impact on the network performance, especially in the presence of RF interference~\cite{baddeley2020impact}. %
\mbox{For example}, the \blefive PHY operating at 2\,Mbps (\blefive 2M) PHY provides higher throughput, but has reduced range and struggles in dense networks due to the lower effectiveness of the capture effect (one needs a considerable power difference to capture the channel with different data). 
In contrast, the \blefive PHY operating at 125\,kbps (\blefive 125K) sacrifices throughput for a longer communication range, but struggles in the presence of RF interference (the longer airtime of packets results in a higher probability of being hit). 

\boldpar{Harnessing the strengths of a PHY at runtime}
As a result, the PHY to be used needs to be carefully chosen and tailored to the specific application scenario at hand. 
This is true under the assumption that the physical layer is \textit{statically} selected at compile time and cannot be changed at runtime -- which is currently the state-of-affairs for all known SF solutions. 
\mbox{In fact}, to date, there has been no attempt to dynamically switch PHY at runtime to leverage on their strengths and increase the reliability and efficiency of communications. 
\mbox{We argue that} one can enable significant performance gains by changing the PHY configuration in individual flooding rounds on demand. 

\begin{table*}[!t]
\caption{Comparison of \NAME with other SF middlewares. \NAME supports the development of multi-PHY protocols -- allowing exploitation of multiple PHYs \textit{in the same protocol}, as well as dynamic selection and (re)configuration of the PHY \textit{at runtime}.}
\vspace{-2.00mm}
\renewcommand\arraystretch{.8}
\scriptsize
\centering
\begin{tabular}{c c c l c c c c}
\toprule
\bf{Framework} & \bf{Public} & \bf{Platform(s)} & \textbf{PHYs} & \bf{Multi-PHY} & \bf{Driver} & \bf{Primitives} & \bf{Protocols} \\
\midrule
A$^2$~\cite{alnahas2017network}                     & \checkmark & TelosB             & IEEE      &            & Polling & Glossy & 2PC, 3PC \\
\midrule
Atomic~\cite{baddeley2019atomic,baddeley2020impact} &            & TelosB, nRF52840   & IEEE, BLE &            & ISR     & RoF    & STA \\
\midrule
Baloo~\cite{jacob2019baloo}                         & \checkmark & TelosB, CC430      & IEEE      &            & Polling & Glossy & Chaos, Crystal, LWB, Sleeping Beauty \\
\midrule
Baloo-nRF52840~\cite{tuchtenhagen2022more}          & \checkmark & nRF52840           & IEEE, BLE &            & Polling & Glossy & \\
\midrule
BlueFlood~\cite{alnahas2019concurrentBLE5}          & \checkmark & nRF52840           & IEEE, BLE &            & Polling & Glossy \\
\midrule
\NAME                                                 & \checkmark & nRF52840           & IEEE, BLE & \checkmark & ISR     & Glossy, RoF & STA, Crystal\\
\bottomrule
\end{tabular}
\vspace{-5.00mm}
\label{tab:osf_comparison}
\end{table*}

\boldpar{Our contributions} 
We hence present Open Synchronous Flooding (\NAME), a multi-PHY SF framework for the Nordic Semiconductor~\texttt{nRF52840} platform. 
In contrast to other public SF middlewares and architectures such as $A^2$~\cite{alnahas2017network}, Baloo~\cite{jacob2019baloo}, and BlueFlood~\cite{alnahas2019concurrentBLE5}, 
which utilize polling-based radio drivers, \NAME implements a novel interrupt-based driver. This allows the complex and time-critical elements of CT-based operations to be handled with relative ease, and provides a robust platform for SF protocol implementation. 
More importantly, the middleware architecture of \NAME enables unprecedented configuration of the flooding rounds: it enables not only fine-tuning of many SF variables (such as the number of transmissions, the maximum number of hops, and the flooding primitive), but also allows the underlying PHY to be switched \textit{at runtime}. 
\mbox{\NAME in fact allows} a dynamic selection of the physical layers supported by the \texttt{nRF52840} (e.g., \ieee or any of the four \blefive PHYs), and therefore supports the implementation of multi-PHY SF protocols able to leverage previous insights on the PHY impact on CT performance~\cite{baddeley2020impact}. 

\fakepar
We make \NAME publicly available\footnote{https://github.com/open-sf/osf.git} as the first SF middleware to support multi-PHY experimentation. 
We run \NAME across data dissemination and collection scenarios on the \dcubee testbed~\cite{schuss2017competition}: our results not only show significant performance improvement under RF interference in comparison to recent studies, but we are the first to benchmark SF for large (118 and 248 byte) data packets.
Furthermore, we address recent findings on the weaknesses of \blefive uncoded PHYs~\cite{baddeley2020impact,jakob20experimental}, where a large number of transmitters can result in a high percentage of collisions, and introduce mechanisms capable of achieving up to 100\% reliability. Finally, we demonstrate how a multi-PHY approach can achieve reliability of the coded PHYs and the latency and energy benefits of the higher data-rate PHYs, delivering near-perfect reliability with energy gains of up to 40\%. We further implement a reactive multi-PHY protocol capable of dynamically adjusting PHY utilization across a time-variant interference scenario, providing $\approx$20\% energy gains without sacrificing reliability.

\boldpar{Paper outline}
We provide background information on how PHY aspects affect SF performance in Sect.~\ref{sec:background}, along with an overview of relevant works in this area. 
Sect.~\ref{sec:design} details architectural aspects of \NAME, and highlights a number of novel mechanisms designed to improve SF performance. 
In Sect.~\ref{sec:results}, we provide an experimental characterization of \NAME's performance across various scenarios, and demonstrate the potential behind multi-PHY solutions w.r.t. the reliability and energy efficiency of communications. 
Finally, we conclude the paper in Sect.~\ref{sec:conclusions} by providing a discussion on how this work may be taken forward in future research.

\section{Background and Related Work}
\label{sec:background}

\boldpar{Enduring popularity of SF} 
Synchronous flooding has been popular in the context of low-power wireless systems since the development of Glossy~\cite{ferrari2011efficient}. 
By exploiting constructive interference and the capture effect, 
this technique can indeed provide highly-reliable flooding for one-to-many communication, and can be also used to achieve fast all-to-all data sharing as shown by Chaos~\cite{landsiedel2013chaos}.
Works such as Codecast~\cite{mohammad18codecast} and Mixer~\cite{herrmann18mixer} provide a more general approach to support many-to-many communication by introducing feedback-driven network coding. 
\mbox{Especially} when combined with channel hopping mechanisms, SF can outperform conventional routing-based solutions in terms of reliability, end-to-end latency, and energy consumption, even in the presence of harsh RF interference~\cite{lim2017competition, ma2020harmony}, as highlighted in the context of the EWSN dependability competition series~\cite{boano17competition,schuss2017competition}. 

\boldpar{Middleware architectures} 
A number of works regard a CT-based flooding round (e.g., Glossy, Chaos, etc.) as primitives atop which SF protocols are built (and are comprehensively covered in a recent survey~~\cite{zimmerling2020synchronous}).
These provide an abstraction which separates the tight-timing-required by CT from in-network information processing.
In this way, with a constant number of CPU cycles between packet receiving and transmitting, any in-network processing operations (e.g., Paxos~\cite{poirot2019paxos}) can be applied. More recently, SF has been integrated into routing-based multi-hop protocols: T-RPL~\cite{istomin2019route} introduces SF to asynchronous CSMA/CA \ieee networks, improving the reliability of the downward traffic in the IETF RPL~\cite{winter2012rpl}, while 6TiSCH++~\cite{baddeley20216tisch++} combines SF and the timeslot-based protocol~\cite{thubert2021architecture}, providing an encapsulation of a CT flood within an \ieee~TSCH slot frame.

However, CT operations rely on low-level control of timers and radio events. As such, SF protocols are notoriously difficult to implement. To address this challenge, Baloo~\cite{jacob2019baloo} introduces a middleware layer providing a well-defined interface for network protocol design; 
specifically focusing on enabling runtime control from the network layer protocol logic over the execution of the
underlying SF primitives. \NAME goes considerably further -- not only supporting SF primitive switching, but allowing full runtime configuration of round variables (e.g., number of transmissions, transmission power) and even the underlying radio PHY.
We compare \NAME against other SF middlewares and architectures in Tab.~\ref{tab:osf_comparison}.

\boldpar{Impact of the physical layer}
A majority of SF research has targeted \ieee devices. Recent studies, however, have verified feasibility over other technologies such as ultra-wideband~\cite{lobba2020concurrent}, LoRa~\cite{liao17lora, bor2016lora, tian2021chirpbox}, and BLE~\cite{baddeley2020impact, nahas2021blueflood, jakob20experimental}. Al Nahas et al.~\cite{nahas2021blueflood} experimentally demonstrated feasibility over \blefive PHYs and proposed BlueFlood, which employs a Glossy-based CT primitive over both \ieee and BLE. Baddeley et al.~\cite{baddeley2020impact} perform large scale investigation of SF over \ieee and the four \blefive PHYs in the presence of radio interference, while also examining how the CT-induced beating effect (sinusoidal patterns of constructive and destructive interference across the packet) can have significant impact on performance depending on the chosen PHY, and providing insights on how this may be countered.

\begin{figure}[t]
	\centering
	\includegraphics[width=.75\columnwidth]{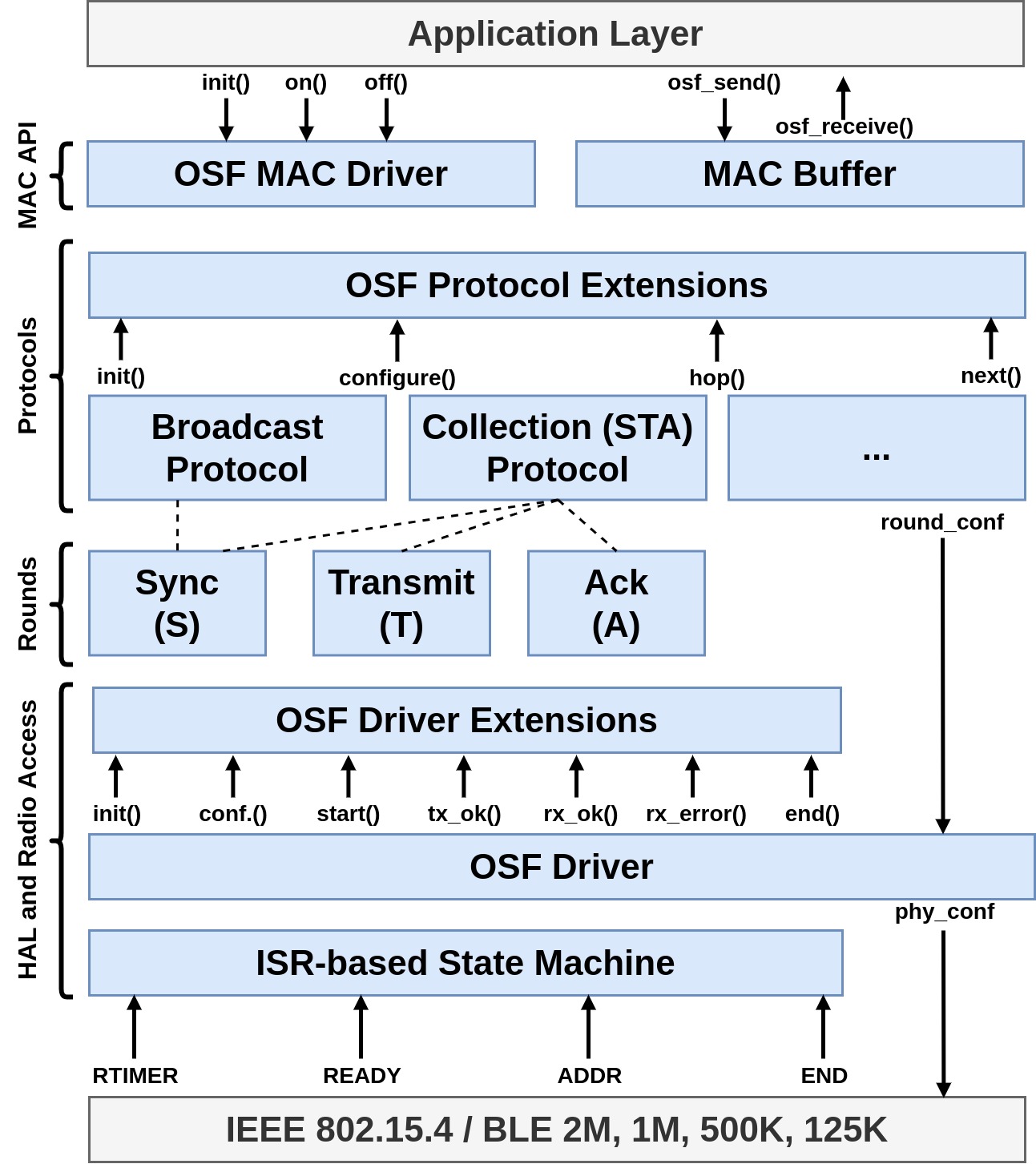}
	\vspace{-3.00mm}
	\caption{Overview of the \NAME stack.}
	\label{fig:osf_stack}
	\vspace{-6.00mm}
\end{figure}

\section{\NAME: Architecture and Implementation}
\label{sec:design}
To the best of our knowledge, \NAME is the first SF middleware supporting multi-PHY protocols. 

\boldpar{\NAME stack}
\NAME employs a middleware approach inspired by recent community efforts, where simple broadcast protocols such as Glossy~\cite{ferrari2011efficient} (which operates in a reception-triggered manner, i.e., \emph{Rx-Tx-Rx-Tx})
and RoF~\cite{lim2017competition} (which operates in a time-triggered manner, i.e., \emph{Rx-Tx-Tx-Tx})
are considered primitives upon which more complex protocols can be built. Fig.~\ref{fig:osf_stack} provides an overview of the \NAME stack. Like other SF middlewares such as Baloo~\cite{jacob2019baloo} and Atomic~\cite{baddeley2019atomic}, \NAME introduces \textit{rounds} as building blocks of network protocol logic. However, \NAME takes this approach significantly further -- not only allowing \textit{runtime configuration} of flooding round parameters (e.g., number of transmissions, transmission power, and maximum number of hops), but also the underlying PHY configuration as handled by the radio HAL. Finally, as a platform for SF experimentation, \NAME introduces a \textit{driver extension} API allowing easy implementation of low-level SF enhancements (such as the RNTX mechanism introduced in Sect.~\ref{sec:results}) without prior knowledge of the underlying radio access.

\begin{figure}[!t]
	\centering
	\includegraphics[width=0.92\columnwidth]{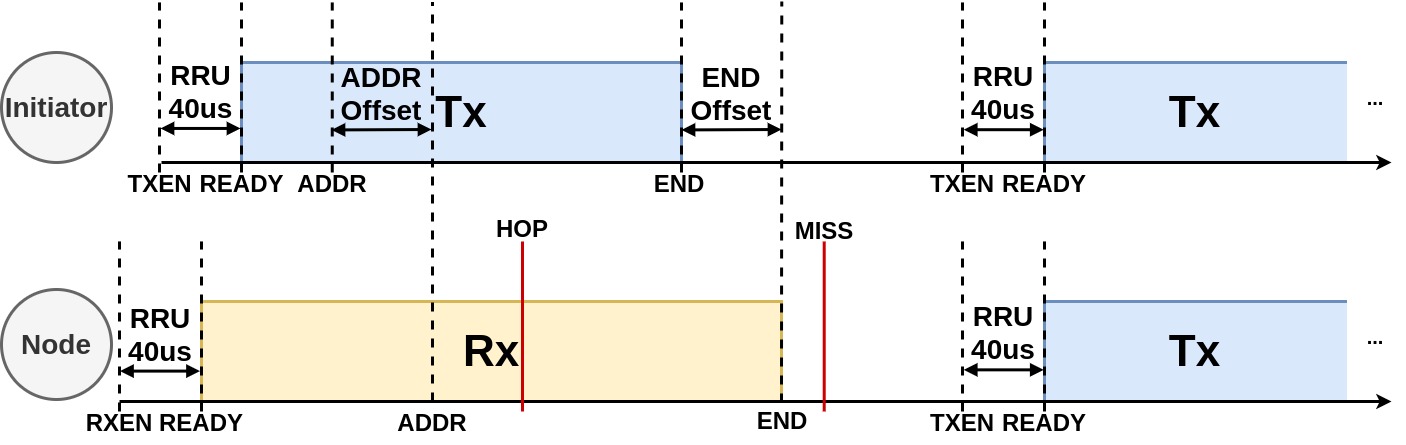}
	\vspace{-3.00mm}
	\caption{Interrupt-based slot management in \NAME.}
	\label{fig:osf_isr}
	\vspace{-2.00mm}
\end{figure}

\boldpar{Interrupt-based driver and slot management}
Unlike many recent SF driver implementations, \NAME uses event-triggered timer and radio interrupts generated by the modern Nordic Semiconductor \texttt{nRF52840} low-power wireless chipset to exercise fine-grained control over a CT-based flood. Fig.~\ref{fig:osf_isr} highlights how, after a 40$\mu$s radio rampup (RRU), the \texttt{ADDRESS}, \texttt{READY}, and \texttt{END} events are used to manage the ISR-based state machine -- where actions such as channel hopping (\texttt{HOP} - approx. halfway between \texttt{ADDRESS} and \texttt{END}) and edge-case errors such as an \texttt{END} \texttt{MISS} can be handled through timer interrupts triggered after these events.
Specifically, \NAME's interrupt-based approach means that the time-critical aspects of SF protocol development are automatically handled by the interrupt service routine (ISR), and can be `ignored' while performing other tasks, while a polling-based approach introduces blocking. This has significant implications when considering USB communication to a host device: rather than scheduling a dedicated host communication time period before or after the flooding, \NAME checks for late timer events and will automatically reschedule itself to account for delays, allowing OSF to run unobtrusively alongside other processes. 

\begin{figure}[!t]
	\centering
	\includegraphics[width=0.95\columnwidth]{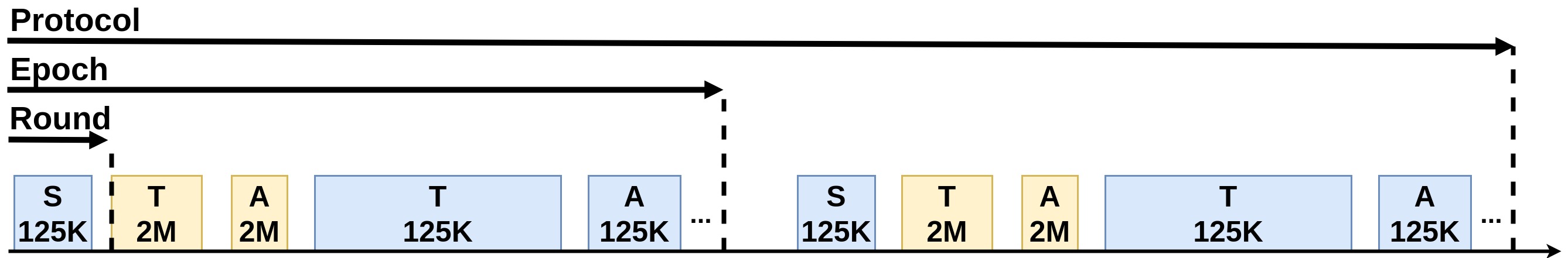}
	\vspace{-2.75mm}
	\caption{PHYs may be configured per-round at runtime.}
	\label{fig:osf_mphy}
	\vspace{-5.00mm}
\end{figure}

\boldpar{Making use of \NAME}
\NAME introduces two features for SF experimentation. Firstly, a novel \textit{driver extension framework} (as shown in Fig.~\ref{fig:osf_stack}) provides hook-ins to key functions within the \NAME driver, as well as \NAME protocols. For example, the RNTX mechanism introduced in Sect.~\ref{sec:results} is implemented as a \emph{driver extension} and introduces a random number of transmissions within each round without touching the base driver, while the backoff mechanism (also introduced in Sect.~\ref{sec:results}) is implemented as a \emph{protocol extension} without modification to the underlying collection protocol.
Secondly, \NAME enables \textit{unprecedented runtime customization} of the SF round configuration and underlying radio PHY.
Fig.~\ref{fig:osf_mphy} demonstrates how \NAME is capable of switching PHY configuration on a per-round basis to build multi-PHY protocols. In this simple example, the \textit{synchronization} (S) round is sent on the PHY with the furthest range, while \textit{transmission and acknowledge} (TA) pairs are allocated either the \blefive\,2M or \blefive\,125K PHY. 
This ensures synchronization of far-away nodes through use of a long-range \blefive\,125K PHY during the S round, while equally providing opportunity for those nodes to transmit back to the sink during the \blefive\,125K TA pair -- all this while nearer nodes may utilize the higher data-rate \blefive\,2M PHY, thus spending less time on-air and saving energy. 


\begin{figure*}[!t]
	\centering
	\begin{subfigure}[t]{0.46\textwidth}
    	\centering
    	\includegraphics[width=\columnwidth]{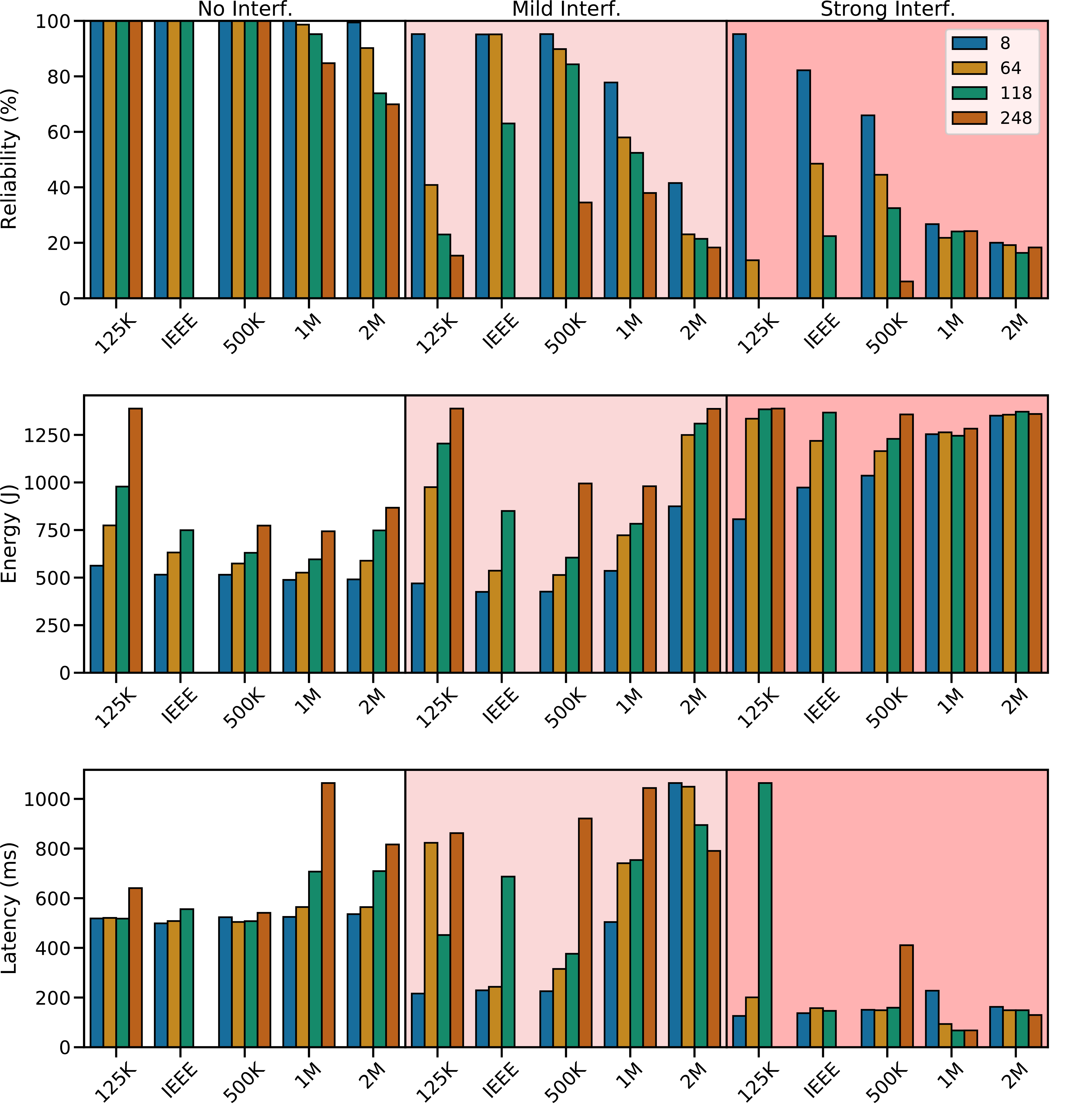}
    	\caption{Data dissemination}
    	\label{fig:bench_dissemination}
    \end{subfigure}	
    \begin{subfigure}[t]{0.46\textwidth}
    	\centering
    	\includegraphics[width=\columnwidth]{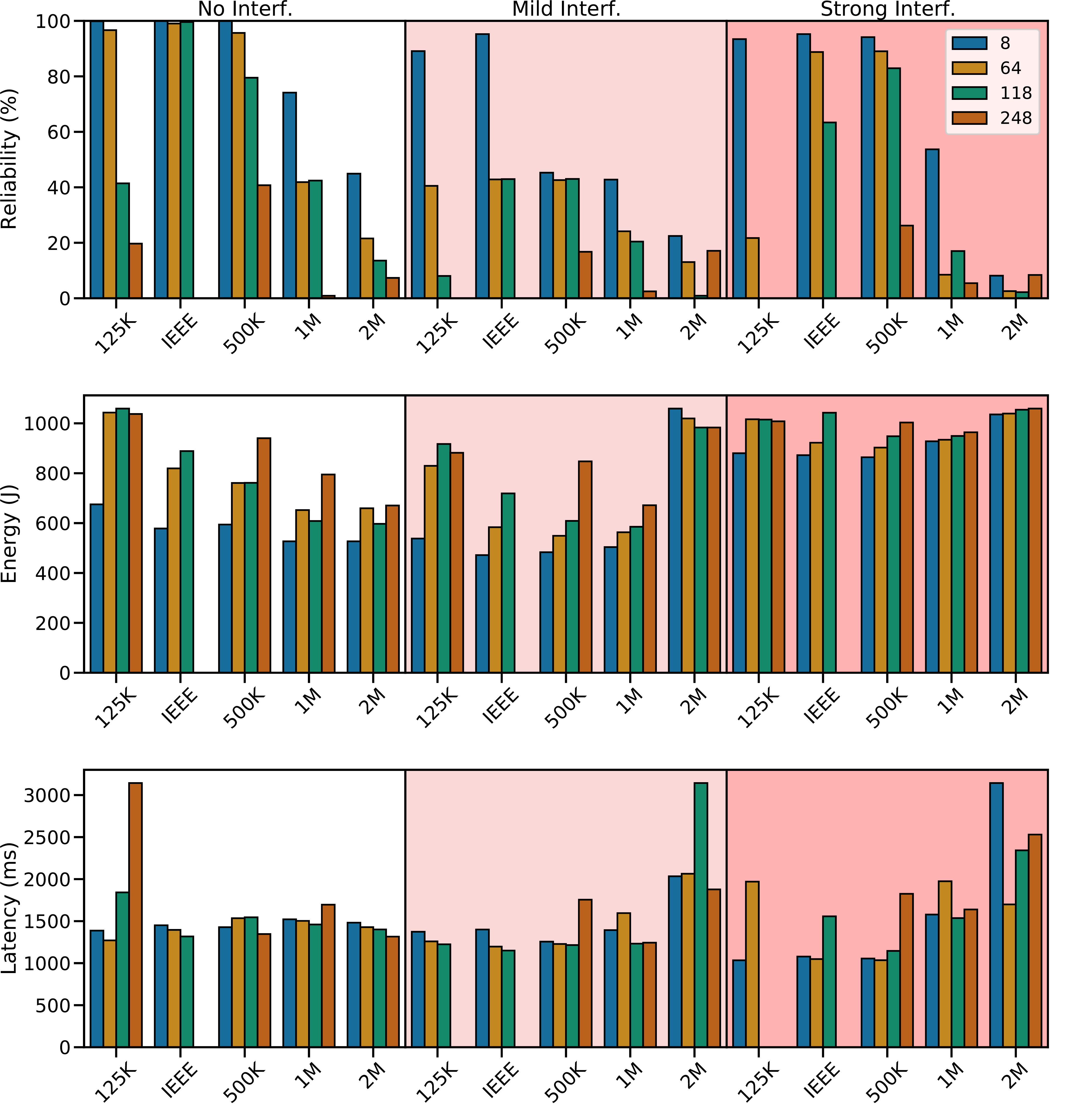}
    	\caption{Data collection}
    	\label{fig:bench_collection}
    \end{subfigure}
    \vspace{-3.25mm}
    \caption{\NAME benchmarking over the RoF primitive for 8, 64, 118, and 248 byte data packets.}
    \vspace{-4.50mm}
    \label{fig:benchmarking}
\end{figure*}

\section{Experimental Evaluation}
\label{sec:results}
We evaluate \NAME in comparison to recent SF benchmarking of a polling-based driver~\cite{baddeley2020impact} based on BlueFlood~\cite{alnahas2019concurrentBLE5}, and provide the first evaluation of SF protocols for larger packet sizes (118 and 248 bytes). We subsequently show that simple randomization and backoff mechanisms are capable of avoiding the large number of collisions occurring when using \blefive uncoded PHYs in dense scenarios~\cite{baddeley2020impact,jakob20experimental}.
Finally, we provide initial data on multi-PHY SF protocols, where we vary utilization between long-range (\blefive\,125K) and high data-rate (\blefive\,2M) PHYs and demonstrate a na\"ive multi-PHY protocol capable of adapting to time-variant interference.

\boldpar{Experimental setup} 
We perform all our experiments on the \dcubee testbed embedding 48 \texttt{nRF52840} nodes. 
We set the transmission power for all nodes to 0\,dBm, the number of transmissions after the first reception ($NTX$) to 6, and the maximum number of slots in any flooding round ($NSLOTS$) to 12. 
RoF~\cite{lim2017competition} with timeslot-level channel hopping is used as the underlying SF primitive for both dissemination and collection scenarios, where the epoch periodicity is set to 200\,ms and 1\,s, respectively. 
For data collection, we implement a modified Crystal~\cite{istomin2016data} protocol (operating over RoF rather than Glossy), where we set the number of transmission-acknowledge (TA) pairs after the initial synchronization (S) round ($NTA$) to 12, and the number of empty TA pairs before exit to 4. Unless otherwise stated, all experiments used \dcubee's 30s aperiodic data generation. These configuration settings are in line with recent benchmarking studies~\cite{baddeley2020impact}. 

\boldpar{Data dissemination and collection using \NAME}
We benchmark the performance of \NAME data dissemination and collection by comparing all \ieee and \blefive PHYs across \textit{no}, \textit{mild}, and \textit{strong} interference levels 
(jamming levels 0, 1, and 3 in \dcubee~\cite{schuss19jamlabng}, respectively). 
Dissemination was evaluated over a \textit{one-to-all} broadcast layout (layout 4 on \dcubee) with a single source transmitting to 47 nodes, while collection was evaluated over an \textit{all-to-one} layout with all nodes transmitting to a single destination (layout 3 on \dcubee).

Fig.~\ref{fig:benchmarking} shows that \NAME exhibits a dependable performance. 
Specifically, in Fig.~\ref{fig:bench_dissemination} dissemination reliability is roughly in-line with similar experiments performed on a polling-based driver~\cite[Fig. 10]{baddeley2020impact}, while both latency and energy are reduced by up to 50\%. Furthermore, in Fig.~\ref{fig:bench_collection} \NAME demonstrates high reliability in collection scenarios for smaller packets across the coded PHYs (particularly under \textit{strong} interference) while retaining energy-efficient operations. 

Additionally, this is the first study to evaluate the impact of larger packet sizes. In addition to 8 and 64 byte data, we evaluate 118 (the \ieee MTU minus the CRC + \NAME header) and 248 bytes data (the \blefive MTU minus the \NAME header). Interestingly, we find that while previous studies suggest that \ieee and \blefive\,500K had similar performance under interference~\cite{baddeley2020impact}, \blefive\,500K is better able to handle the 118B packets, likely due to the reduced on-air time of the PHY. However, when considering 248B (too large for \ieee) we see considerable performance degradation across all PHYs when under any sort of interference.

\boldpar{Random NTX and random backoff}
While higher data-rate PHYs consume less energy, due to the beating effect and desynchronization \blefive uncoded PHYs are more susceptible to packet errors as the number of simultaneous transmitting nodes increases -- meaning that the reliability of SF protocols adopting these PHYs drops significantly even when under no external interference~\cite{baddeley2020impact}. 
\NAME introduces two mechanisms to reduce the number of simultaneous transmitters on uncoded PHYs, thereby improving reception rate: (i)~a random number of transmissions after the first reception (RNTX) that can be applied to any SF primitive; (ii)~a fixed-threshold random backoff on the `T' round (BACKOFF) specifically designed for SF-based data collection protocols.

\italicpar{RNTX} 
Before each round each node calculates a random number on the interval [$NTX_O$, 2$NTX_O$], with $NTX_O$ representing the given NTX. This number is then applied to the round configuration, ensuring that transmission slots towards the end of the round will experience fewer numbers of transmissions as nodes will have random NTX. This approach can therefore provide considerable gains in dense network areas.

\italicpar{Random backoff}
While protocols such as Crystal~\cite{istomin2016data} have previously relied on significant capture effect to allow neighboring nodes to concurrently transmit different data during the same flooding round, recent studies have shown that \blefive uncoded PHYs perform poorly in this manner~\cite{baddeley2020impact,jakob20experimental}.
We implement a simple random backoff mechanism to reduce the number of initiators within different-data transmission (T) rounds. Source nodes initiate a flood with a fixed probability of 80\%; otherwise, they act as a forwarder for this round.

\begin{figure}[!t]
	\centering
	\includegraphics[width=.7\columnwidth]{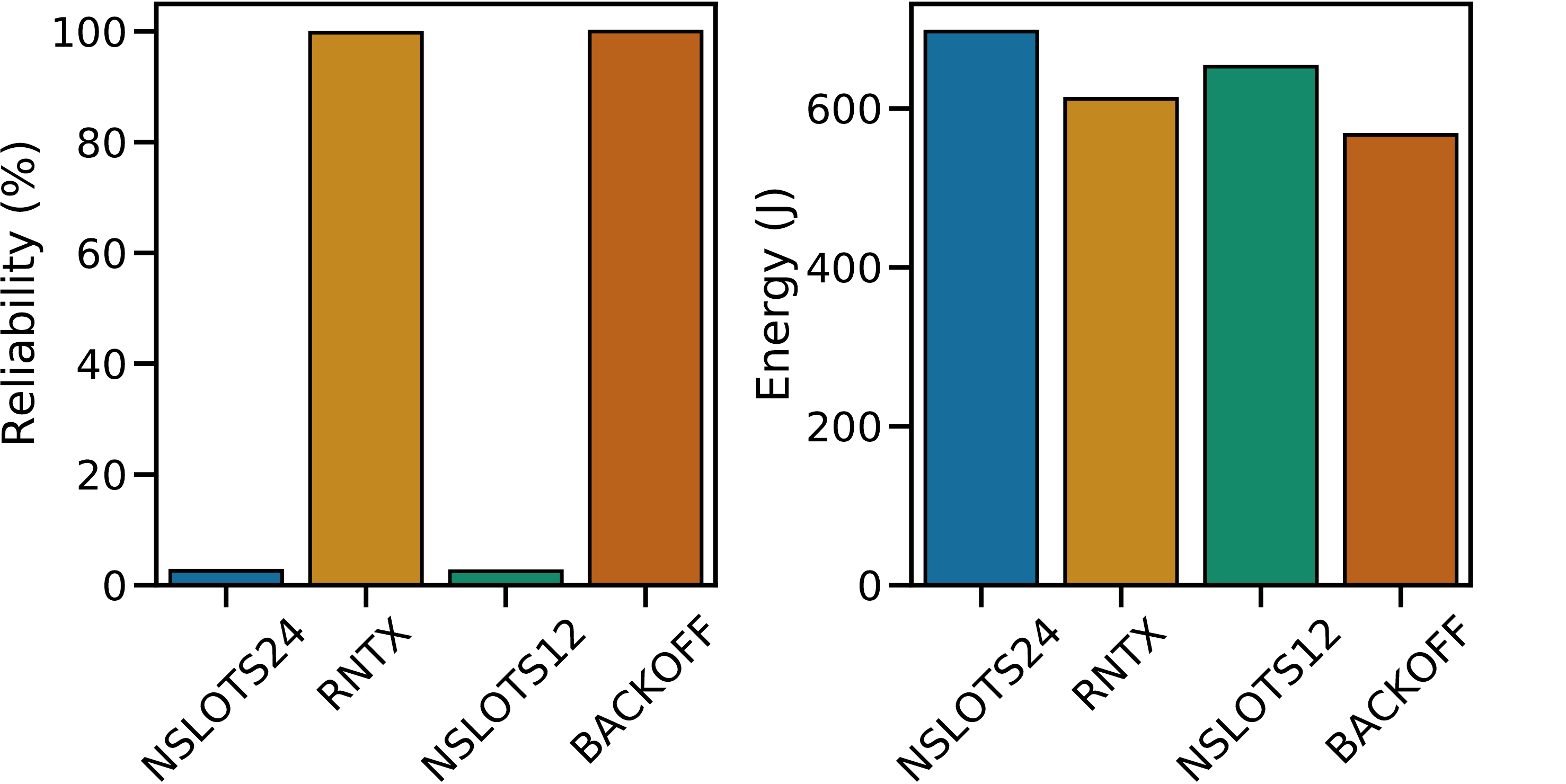}
	\vspace{-2.50mm}
	\caption{RNTX and backoff address collisions in \blefive\,2M.}
	\label{fig:rntx_backoff}
	\vspace{-4.50mm}
\end{figure}

\italicpar{Results}
To evaluate these mechanisms, we set the PHY to \blefive 2M, and run experiments on \dcubee's dense layout (19 source nodes), where all source nodes are roughly one hop away from the sink. To ensure a high degree of competition between the nodes, 5\,s periodic data generation is used, while the data length is set to 64 bytes. To ensure sufficient slots towards the end of the packet, the maximum number of slots for the RNTX mechanism is set to 24 (i.e., each node will transmit between 6 and 24 times), while the maximum number of slots remains at 12 for the random backoff mechanism.
Fig.~\ref{fig:rntx_backoff} shows that regardless of whether the maximum slots is set to 12 (\texttt{NSLOTS12}) or 24 (\texttt{NSLOTS24}), \blefive\,2M has extremely poor reliability. However, when using \textit{either} RNTX \textit{or} the random backoff, 100\% reliability can be achieved, while the backoff mechanism consumes less energy than RNTX.

\begin{figure}[!t]
	\centering
	\includegraphics[width=\columnwidth]{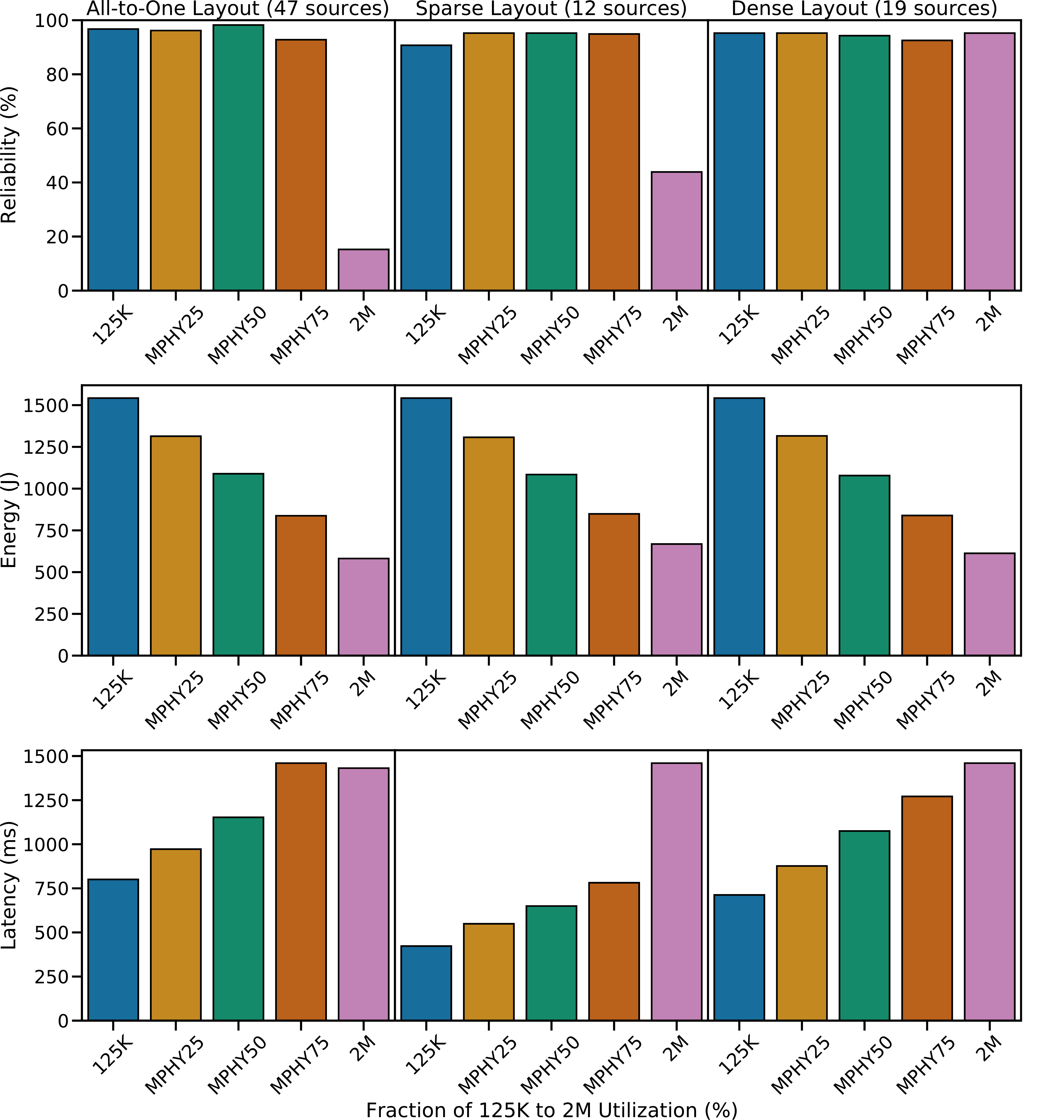}
	\vspace{-6.50mm}
	\caption{Static multi-PHY patterns (no interference).}
	\vspace{-1.50mm}
	\label{fig:osf_results_mphy}
\end{figure}

\boldpar{Static multi-PHY benchmarking}
We evaluate \NAME's multi-PHY capabilities of over 3 collection scenarios: an \textit{all-to-one} layout with 47 sources transmitting to a single node near the topology center; a \textit{sparse} layout with 12 sources (mostly located at the edge of the network with some in the middle) and a destination (located in one corner); finally, a \textit{dense} layout with 19 sources and the destination all in the same room. Fig.~\ref{fig:osf_results_mphy} shows benchmarking of static multi-PHY patterns with increasing \blefive\,2M utilization of the collection protocol TA pair schedule -- from 0\% (i.e., all \blefive\,125K) to 100\% (i.e., all \blefive\,2M) at 25\% increments. Specifically, after an initial S round at \blefive\,125K, static \blefive\,125K/2M patterns
are repeated across the epoch's 12 TA pairs.
For a single-PHY approach, while the \blefive\,125K is capable of achieving near 100\% reliability, the high data-rate \blefive\,2M PHY is unable to successfully deliver all messages (please note that this experiment does \textit{not} make use of the RNTX or random backoff mechanisms previously introduced). However, across all layout scenarios, a multi-PHY compromise maintains a high reliability while introducing energy gains of $\approx$40\% with \texttt{MPHY75} (i.e., where one of every four TA pairs is configured with the \blefive\,125K PHY, while the rest are configured with the \blefive\,2M PHY). 

\begin{figure}[!t]
	\centering
	\includegraphics[width=0.95\columnwidth]{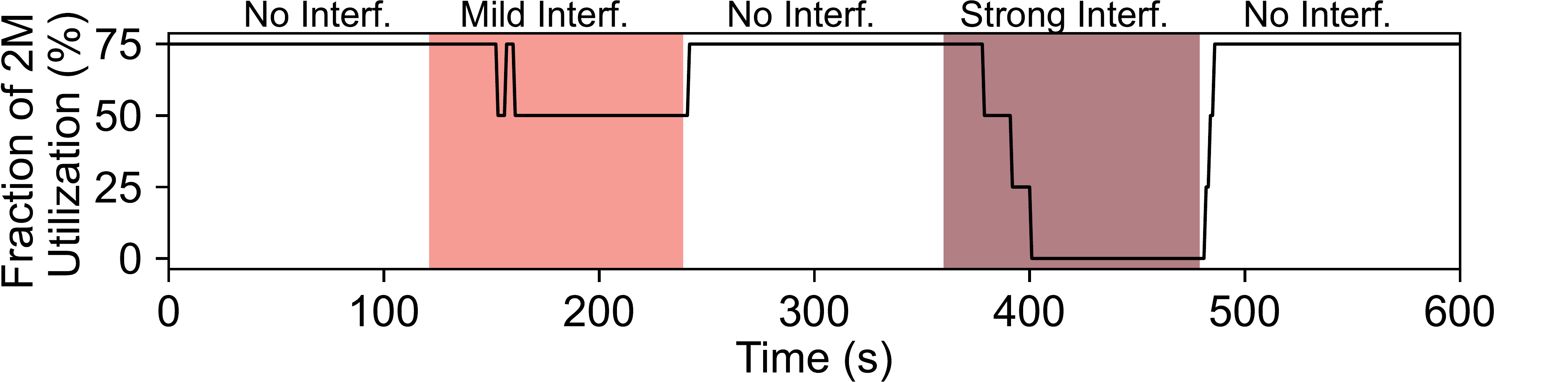}
	\vspace{-3.00mm}
	\caption{Timeline of 2M utilization in Fig.\,\ref{fig:osf_mphy_interference} sparse layout.}
	\label{fig:mphy_pattern}
	\vspace{-3.50mm}
\end{figure}

\boldpar{Dynamic multi-PHY protocol}
Benchmarking of static PHY patterns in Fig.~\ref{fig:osf_results_mphy} fails to account for diverse networks, or changing RF conditions. 
That is, such an approach needs to be configured at compile-time and is unsuited to real-world scenarios. 
We therefore implement a na{\"i}ve dynamic multi-PHY protocol, based on the receive ratio at the sink, as an initial demonstrator of \texttt{OSF}'s multi-PHY runtime capabilities and their potential. Specifically, we devise a time-varying interference scenario as per Fig.~\ref{fig:mphy_pattern}, where levels move from \textit{no}, \textit{mild}, \textit{no}, \textit{strong}, and finally returning to \textit{no interference}. 
 
\noindent
In such a scenario, uncoded PHYs perform well during the periods of \textit{no interference}, yet struggle during the interference periods. On the other hand, \blefive\,500K can perform well even under \textit{strong interference} (unlike BLE\,125K~\cite{baddeley2020impact}). We therefore utilize a \blefive\,500K/2M split rather than the 125K/2M split previously employed in the static benchmarking (under no interference). 
In every epoch, the destination node calculates the percentage of \textit{late} nodes (i.e., sources that have not been received within the 30s aperiodic time window). Based on this percentage, an appropriate multi-PHY pattern is selected according to Tab.~\ref{tab:pattern_selection} and disseminated to the network. 

\begin{figure}[!t]
	\centering
	\includegraphics[width=0.95\columnwidth]{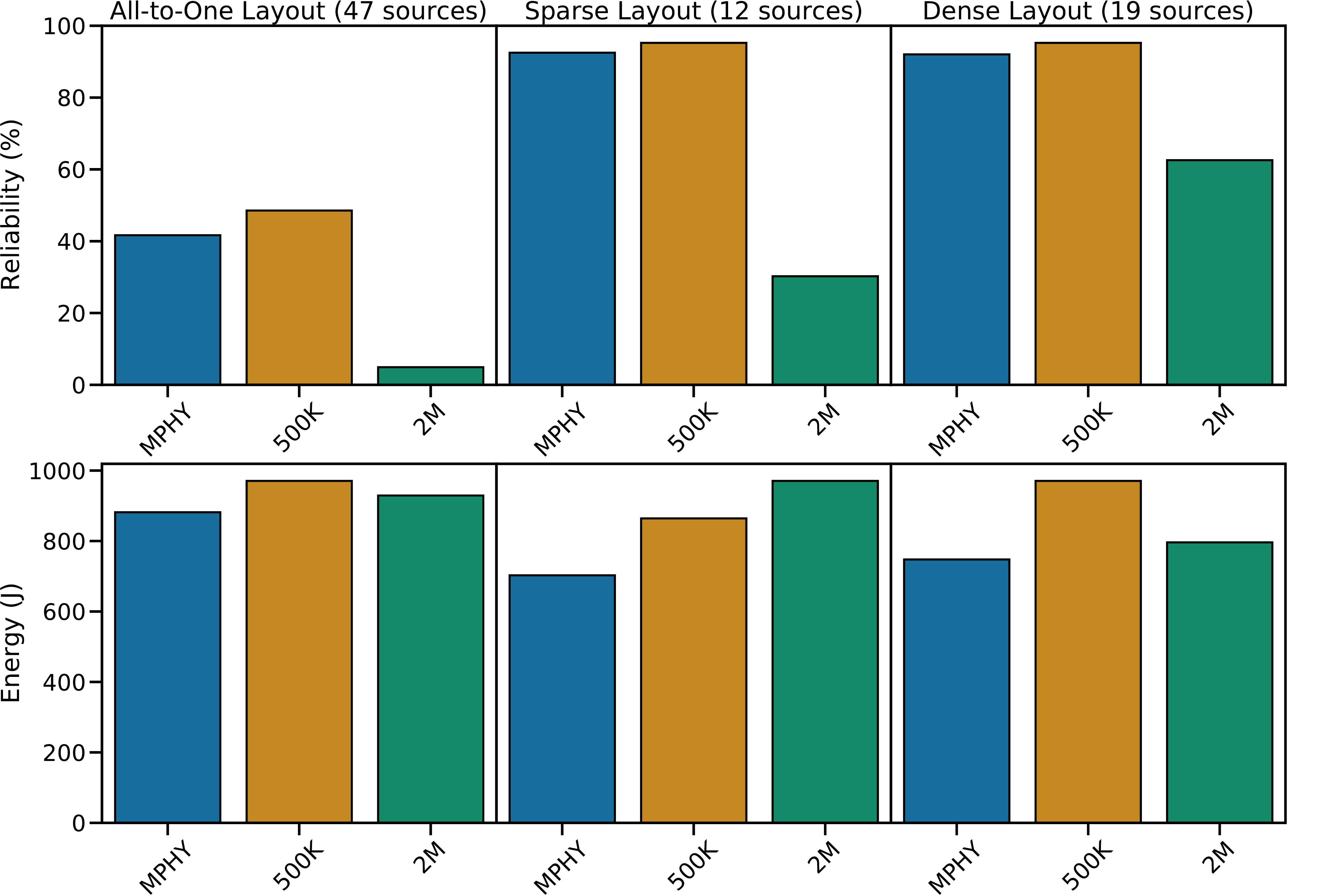}
	\vspace{-2.75mm}
	\caption{Performance of a na{\"i}ve multi-PHY protocol over time-variant interference compared to single-PHY settings.}
	\label{fig:osf_mphy_interference}
	\vspace{-4.50mm}
\end{figure}

\noindent
Fig.~\ref{fig:osf_mphy_interference} compares this na{\"i}ve multi-PHY approach against \blefive\,500K and \blefive\,2M single-PHY across the three \dcubee layouts previously outlined in Fig.~\ref{fig:osf_results_mphy}. Due to interference, \blefive\,2M is incapable of achieving high reliability on any of the layouts, while \blefive\,500K is able to escape -- providing close to 100\% reliability in the sparse and dense layouts. As expected, the multi-PHY approach is able to provide similar levels of reliability to \blefive\,500K, with significantly reduced energy ($\approx$20\%). Furthermore, Fig.~\ref{fig:mphy_pattern} demonstrates how the multi-PHY protocol adapts during the experiments, dynamically switching to greater \blefive\,500K utilization during periods of interference, while increasing \blefive\,2M utilization during periods without interference.


\section{Discussion and Outlook}
\label{sec:conclusions}
This paper has introduced \NAME, a novel SF architecture and middleware supporting \textit{runtime configuration} of the underlying PHY. While some recent works have leveraged the multi-PHY support of modern low-power wireless chipsets such as the \texttt{nRF52840} to support routing-based \ieee solutions with SF data dissemination~\cite{istomin2019route, baddeley20216tisch++}, it has been shown that the underlying physical layer phenomena (namely capture effect, beating effect, and desynchronization) play a significant role in the overall performance of SF protocols~\cite{baddeley20216tisch++}. These findings suggest that specific PHYs may enhance, or hinder, SF performance within certain scenarios -- providing an argument for the research community to research and develop new SF protocols with multi-PHY capabilities. 

\begin{table}[!t]
\caption{Multi-PHY pattern selection.}
\vspace{-3.00mm}
\renewcommand\arraystretch{.9}
\scriptsize
\centering
\begin{tabular}{c c c c c}
\midrule
\# of late nodes & $[0\%, 25\%]$ & $(25\%, 50\%]$ & $(50\%, 75\%]$  & $(75\%, 100\%]$ \\
\midrule
2M utilization & MPHY75 & MPHY50 & MPHY25 & MPHY0\\
\bottomrule
\end{tabular}
\vspace{-1.00mm}
\label{tab:pattern_selection}
\end{table}

To this end, we have made \NAME publicly available as a platform for further research. In support of this, we have benchmarked \NAME across the \dcubee testbed for single-PHY dissemination and collection scenarios, where \NAME demonstrates notable performance improvements under interference when compared to polling-based SF implementations, and experiments on larger packet sizes suggest issues when using the full \blefive MTU. Furthermore, the \NAME middleware not only features network-level SF protocol construction (as with other SF middlewares~\cite{alnahas2017network,baddeley2019atomic,jacob2019baloo}), but provides unprecedented runtime configuration options, as well as a low-level driver extension framework to allow development of enhancements to the \NAME driver while separating the complexities of radio access and the interrupt-based state machine. Based on this, we have implemented and evaluated two simple mechanisms that are capable of overcoming the critical weaknesses of uncoded \blefive PHYs in dense environments -- achieving 100\% reliability as opposed to near-total collisions.

Fundamentally, this paper has validated \NAME's multi-PHY approach through evaluation on the \dcubee testbed, where we show that static PHY patterns (of varying utilization between \blefive\,125K and 2M) achieve high reliability whilst reducing energy and latency. Furthermore, we have demonstrated the effectiveness of even a na{\"i}ve approach to dynamic runtime PHY switching, and achieve considerable energy gains over that of a single-PHY protocol when taking into account time-variant interference scenarios. While the majority of SF research has, to date, only considered controlled environments, these early results suggest that an intelligent multi-PHY approach can significantly improve SF protocol performance for unpredictable real-world scenarios.


\bibliographystyle{IEEEtran}
\bibliography{references}

\end{document}